# Using word embedding for environmental violation analysis：Evidence from Pennsylvania unconventional oil and gas compliance reports

Dan Bi[a], Erlong Zhao[a], Ju-e Guo[a], Shaolong Sun[a,*], Shouyang Wang[b,c,d]

[a]School of Management, Xi'an Jiaotong University, Xi'an 710049, China
[b]Academy of Mathematics and Systems Science, Chinese Academy of Sciences, Beijing 100190, China
[c]School of Economics and Management, University of Chinese Academy of Sciences, Beijing 100190, China
[d]Center for Forecasting Science, Chinese Academy of Sciences, Beijing 100190, China
*Corresponding author. School of Management, Xi'an Jiaotong University, Xi'an 710049, China.
Tel.: +86 15911056725; fax: +86 29 82665049.
E-mail address: sunshaolong@xjtu.edu.cn (S. L. Sun).

**Abstract**

With the booming of the unconventional oil and gas (UOG) industry, its inevitable damage to the environment and human health have attracted the public more and more attention. The data source of this article is from the Department of Environmental Protection (DEP) in Pennsylvania, USA, we used the compliance reports of UOG from 2008/1/1 to 2018/12/31. The purpose of this article is to use text mining and natural language processing techniques to analyze the text data of UOG violations, which are difficult to handle manually, and discover the general rules of violations. First of all, this article makes a statistical analysis of the violation codes, which are divided into "Administrative" and "Environmental Health & Safety" violations. Afterward, based on the compliance reports of "Environmental Health & Safety" violations, we establish the original corpus. Thirdly, based on the previous literature as well as the statistical analysis of violation codes, we traverse the original corpus to select the keywords of UOG violations: "Contaminant", "Location" and "Operation". After text preprocessing, the corpus is put into the Skip-Gram model to train word embedding, then the keywords similarity analysis is performed. At last, we discuss the general rules and internal mechanisms of environmental violations and providing the supervision suggestions. Overall, this research sheds light on the UOG on-site production supervision and regular patterns of production violations both for government and practitioners.

# 1. Introduction

Unconventional oil and gas (UOG) have become one of the most important sources of energy reserves and fossil fuels worldwide, the rich reserves and wide range of uses have made it greatly valued. However, due to the special reservoir structure of UOG, some extraction technologies may cause irreversible damage to the environment and human health, such as geological structures damage, groundwater pollution and earthquakes (Casey et al., 2019; Sun et al., 2019; Torres et al., 2016; Werner et al., 2015). Thus there is an urgent need for comprehensive analysis about violation regular pattern and mechanism in UOG life-cycle production.

Articles that probe into the risk and violations associated with UOG life-cycle production can be divided into mainly three categories: environment influence research, geological structure impact, and human health threat. Rahm et al., (2015) analyzed the shale gas violations association relationship in Marcellus using Ordinary least squares (OLS) and One-way Analysis of Variance (ANOVA) test. Werner et al., (2015) examined the impact of unconventional gas development associated with environmental health by a comprehensive literature review from 1995 to 2014. However, a large number of UOG violation accident data is unstructured data, most of the current literature only use traditional quantitative tools after a lot of manual statistics process.

Marcellus shale gas reservoir is one of the most productive and notable UOG reservoirs in Pennsylvania state, the USA. And Pennsylvania state government-owned the environment regulation autonomy exerted by the Pennsylvania Department of Environmental Protection (DEP). Pennsylvania DEP contributed to supervising and managing various environmental issues, such as underground water pollution, air pollution, lands, and nature biosecurity. To track and analyze the environmental impact of UOG pollution accidents, it has established a particular compliance report to record the violation operation and relative production parameters. This paper is based on the UOG compliance report from Pennsylvania DEP to make a thorough inquiry of the mechanism behind the violation incidents.

We focus on the environmental and health violation regular pattern by exploring UOG production compliance reports published by Pennsylvania DEP from 2008/1/1 to 2018/12/31. First of all, we select the key factors of violation based on violation code analysis and previous literature review. And then we use the word embedding model Skip-Gram, to convert the "Environmental Health & Safety" type of "Inspection Comment" into structured data. Finally, we analyze the regular pattern and mechanism of UOG environmental and health safety violations based on similarity analysis. The framework of this paper is shown in **Figure 1**.

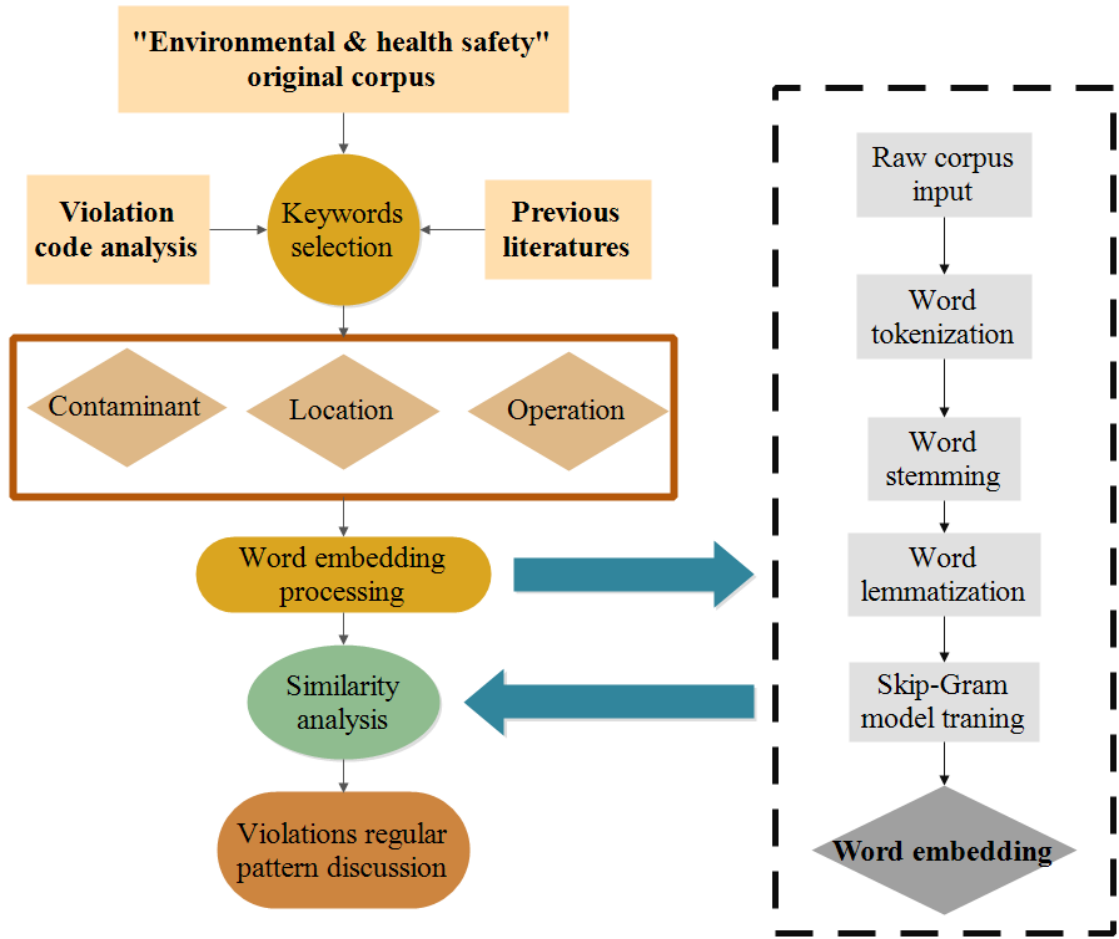

**Figure 1.** The framework of UOG violation analysis

To reach our goal, the following questions should be addressed in this paper:

1. What can we learn from the operator's comment observation description?

2. What causes different types of violations?

3. What are the regular patterns and internal mechanisms of environmental violations?

The contributions of this paper are as follows: (1) this is the first one (to our best knowledge) using nature language processing into UOG violation analysis; (2) we make full use of text data to excavate the mechanism in this kind of violations; (3) our work shed light on the UOG on-site supervision and regular patterns of production violations both for government and practitioners.

The structure of this article is as follows: **Section 2** discusses the literature on UOG production and violation accidents; **Section 3** states the results of statistical analysis of violation codes; **Section 4** will describe the criteria and methodology used in this article; **Section 5** carries out keywords selection and similarity analysis. After discussion, the general law of UOG violations is obtained; **Section 6** summarizes the results, limitations, and future work of this research.

# 2. Literature review

In this section, we discussed the general concept of UOG life-cycle production stages, the attempts to analyze its environmental health impact, and the approaches used nature language processing for support decisions. This exhaustive review aims to find pending questions and gaps in this field.

UOG is rapidly expanding globally with high utilization value. The largest shale gas reserves are China, Argentina, Algeria, the USA, and Canada (U.S. Energy Information Administration, 2015). The UOG production stages and be described into six stages, which are: (1) Drilling wells, including horizontal and vertical wells; (2) Hydraulic fracking and proponent injection; (3) Flowback recovery (initially produced water); (4) Oil and gas flow out of the well; (5) Storage and pipeline transportation; (6) Abandonment wells (Torres et al., 2016).

A lot of research work is also thinking about whether the successful shale gas extraction project from the United States can be copied by other countries, as well as some important historical inspirations. Castro-Alvarez et al. (2018) used the USA as a case study to present comprehensive insights from the environment, community, and economic impact during UOG development, to shed the light on Mexico UOG development for government regulation enacting. Davis and Sims (2019) evaluated the USA UOG industry from the perspective of market operations, and the empirical and simulation confirmed the importance of technological advances and the price of nature gas. Wang and Jiang (2019) provides a novel forecasting model framework for forecasting shale gas production in Pennsylvania and Texas. Harleman and Weber (2017) simulate different scenarios to contract the government regulation of UOG development in the UK and the USA, the result showed that the superiority of UOG policy in the USA. However, Solarin et al. (2020) tested and analyzed the sustainable performance of shale gas in the United States from 2000 to 2018. The analysis results showed that there was a lag in its production. The shale gas prediction model should add more policy considerations.

The contaminate components that hurt the environment and human health mainly come from the stages of hydraulic, proponent injection and flowback recovery. The causes of these violation incidents both came from human administrative and technical error as well as inevitable technical consequences. Empirical research on environmental violation reports lunched by DEP is also increasing. Chen and Carter (2016) surveyed the amount of fresh water used in unconventional fracturing wells in the USA from 2008 to 2014, and the findings help the industry with its water management plans. Butkovskyi et al. (2019) conducted a study on fracking water demand and wastewater management in the Dutch Posidonia Shale, which established a complete research framework and the results could contribute to future legislative work in this area. Hernández-Espriú et al. (2019) simulated four hydraulic fracturing water use scenarios to evaluate the water pressure indicators of different blocks and designed a shale gas production plan, which can evaluate water restrictions in unconventional industries. Centner and O'Connell (2014) reckoned that there is still room for the USA to improve UOG environmental regulations, he points out some deficiencies of both state governments and enterprises and the direction for improvement. Deziel et al. (2020)

reviewed the literature about human health associated with exposure to UOG development. Maloney et al. (2017) aimed at the risk of UOG wells spills in the USA and conducted a statistical analysis of a total of 6622 reported spills to guide regulations. Sun et al. (2019) investigated the trace elements of fracking wastewater and flowback then developed a water management strategy. Werner et al. (2015) concerned about the hazards caused by UOG development, which can be harmful to human health, and the review results revealed that there are still lacking sufficient evidence to rule out possible health impacts.

With the rising of sustainability and green industry calling, the self-discipline of enterprises is not enough, the government must exert strong environmental regulation. A survey about the forest fragmentation affected by Marcellus shale gas is conducted by Langlois et al. (2017), the results alerted the government and practitioners about the balance between nature and industry flourish. Oke et al. (2020) considers shale gas production and water demand scheduling under uncertain conditions. The results of this study can provide companies with plans to increase expected profits and save fresh water. Wang and Zhan (2019) used the sustainability evaluation model to conduct empirical analysis on the shale gas industry in Chongqing and Sichuan, China, and obtained important factors affecting sustainable development. Guo et al. (2019) analyzed the regulation in Pennsylvania shale gas development using Hierarchical linear modeling, and the result confirmed the positive effect of strict regulation and economical punishment to decrease the number of violations. Centner (2016) summarized five points to contain environmental health trouble during UOG development in the USA, it also has a certain enlightening effect on the formulation of policies in other countries. Chen et al. (2020) evaluated the tradeoffs in water and carbon footprints of UOG compared with conventional nature gas in China, and the results indicated that UOG can lead to a less greenhouse gas emission future. Cronshaw and Grafton (2016) from a policy perspective conducted a comprehensive analysis of the USA economic benefits and regulation of UOG. Torres et al. (2016) introduced and analyzed each of the risk assessment tools applied in UOG hydraulic fracturing. Zirogiannis et al. (2016) explored the regulation of different states in the USA and ranking the regulation elements according to the effect on controlling UOG environmental health impact.

In recent years, many scholars have also paid attention to the violation reports issued by the government, hoping to get some inspiration and lessons for reference. Guo et al. (2017) contrasted the USA and China monitoring, reporting, and verification systems, and the result showed that China should pay more attention to water contamination. And then Guo et al. (2019) applied a three-level hierarchical linear model. However, due to the limitation of the unstructured compliance report, the quantitative works greatly depend on the manual preprocessing work (Guo et al., 2017, 2019; Rahm et al., 2015), and the alternative kind of research method is the literature review. Manda et al. (2014) invested the different types of UOG well pads on how to influence the environmental violations and wastewater volumes according to violations reports. Rahm et al. (2015) tried to figure out the key characters causing different types of UOG environmental violations, the results indicated that the environmental violations most likely happened in drilling and hydraulic fracturing. Abualfaraj et al. (2016) analyzed the compliance violations for nature gas from 2000 to 2014 and conducted a logistic regression to test the condition for violation accidents, and the empirical study confirmed that UOG is more likely to occur environmental violations. There is a rising concern about UOG development to human health. Casey et al. (2019) conducted an empirical study to evaluate the mediating mechanisms between the UOG development and birth outcomes from 2009 to 2013, however, the result reveals that there is no obvious association between them.

# 3. Data collection

We collect unconventional **"Oil and Gas Compliance Reports"** from the Pennsylvania Department of Environmental Protection (https://www.dep.pa.gov/). The period is from 2008/1/1 to 2018/12/31. And we choose the "**Inspect Comment**" in the **"Violation type"** of "**Environmental Health & Safety**" as our research objectives to reveal the regular patterns and internal mechanism of UOG violations. Following are elements that are used for Oil and Gas reports in this paper:

**"Oil and Gas Compliance Report"**: it provides a listing of all inspections, violations, and enforcement that occurred.

"**Inspect comment**": These are text comments supplied by the inspector of the inspection. DEP requests that this column be used to note any significant observations during the inspection. Qualifying observations that might appear out of the ordinary to DEP staff reviewing assessment reports, indicating if annular spaces are plumbed to tanks or not accessible, or describing well construction modifications or design clarifications are some examples.

**"Violation type"**: the type of violations (Administrative or Environmental Health & Safety).

**"Violation code & description":** the code and description for the violation that was identified during an inspection.

## 3.1 Violations statistical analysis

There are three categories for violation types, which are "**Administrative**", "**Environmental Health & Safety**" and "**None**" in this paper: (1) "**Administrative**" means this violation is about operators conducting administrative violations, such as delay reporting or not cooperating with the inspectors. (2) "**Environmental Health & Safety**' denotes the violation is harmed to human health and environment, such as flowback water exposure to groundwater and fossil fuel leak. (3) "**None**" represents there is no violation exist. After removing the null value, we got a total of 80546 comments data, for "**None**" (74489), "**Administrative**" (1902), and "**Environmental Health & Safety**" (4155) as shown in **Figure 2**.

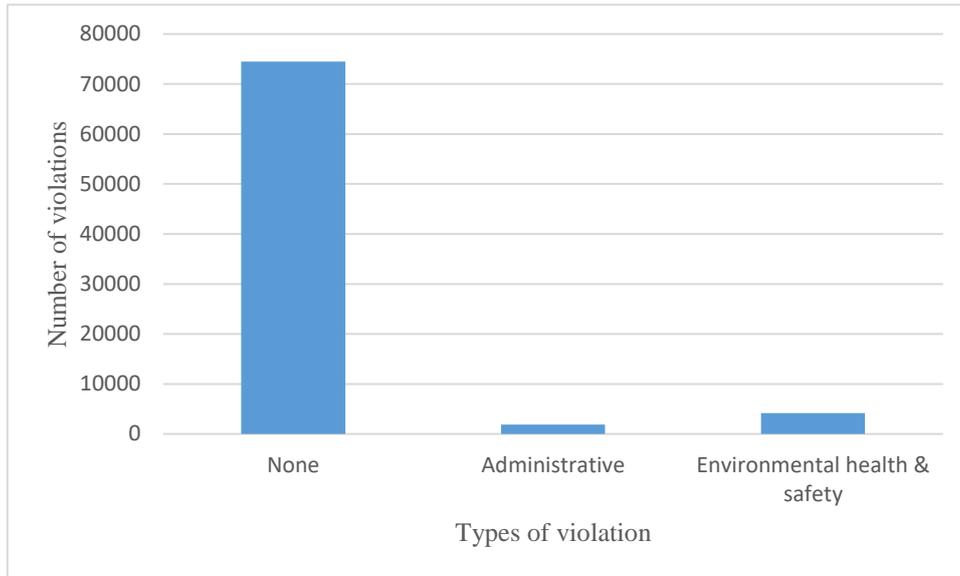

**Figure 2.** The categories of violation type

We make a comprehensive analysis of violation compliances from 2008 to 2018, and the overview is shown in **Figure 3**. As can be seen from the figure, 2010 year (1532), 2011 year (1370) and 2018year (1256) were the years with the most violations. We note that in terms of violations, a total of 9316 violations occurred between 2008 and 2018, but only 6057 violations were selected in this article. This is because we selected violations based on non-empty observation comments, aiming to find the potential regular patterns of violations from the observation records. It can also be seen that a part of the violations did not record the corresponding observations and comments, so they were eliminated.

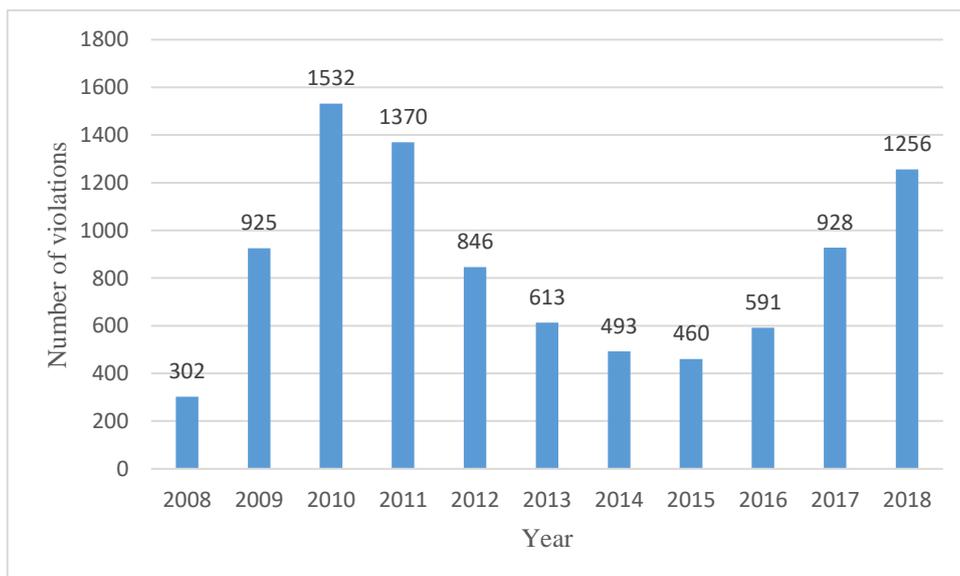

**Figure 3.** The number of violations from 2008 to 2018

There are mainly seven regulations applied to take charge of the UOG industry in PA state, which are regulations of *Unconventional Wells, Oil and Gas Wells, Oil and Gas Conservation, Water Resources, Wastewater Treatment Requirements, Erosion and Sediment Control,* and *Dam Safety.* We counted the most frequent violations in this decade. The top 5 frequent violations are as follows: (1) "Failure to properly store, transport, process or dispose of residual waste." (664 occurrences); (2) "Failure to minimize accelerated erosion, implement E&S plan, maintain E&S controls. Failure to stabilize site until total site restoration under OGA Sec 206(c)(d)" (531 occurrences); (3) "OIL AND GAS Act 223-General. Used only when a specific OIL AND GAS Act code cannot be used" (459 occurrences); (4) "Failure to properly control or dispose of industrial or residual waste to prevent pollution of the waters of the Commonwealth." (321 occurrences); (5) "Discharge of pollution material to waters of Commonwealth." (300 occurrences). First of all, it can be found from the above that the most frequent processes of sewage leakage are in the storage, transportation, and treatment stages. Second, the most frequent violations are wastewater leakage, followed by pipeline erosion and sedimentation.

# 4. Methodology

In this section, we elucidate the framework and methodologies of both machine learning and word embedding as well as the criteria for environmental violations key factors classification. There are three essential steps to form a general concept of UOG life-cycle violation characteristics: (1) Preprocessing the raw data of UOG compliance reports. (2) Preprocessing, which includes word tokenization, removing stopwords, stemming, and lemmatization. (3) Using the Skip-Gram model to train the word embedding.

## 4.1 Criteria for environmental violations

As summarized by Wisen et al. (2019) and Watson and Bachu (2009), three criteria

must be met for environmental violations to occur; these criteria have been adopted in this study, that is:

1. a leak source;
2. a driving force; and
3. a leakage pathway.

In this paper, the "Contaminant" as the leak source, which selected from the original corpus of the "Environmental Health & Safety" inspection comment, represents the materials that caused pollution. While "Operation" corresponds to the driving force, which indicates the behavior and operation processed during UOG production. "Location" implicates the leakage pathway, this kind of words can fully represent the place where the violation occurred or where be polluted.

## 4.2 Text preprocessing

In engineering practice, the original data will have missing values, duplicate values, etc., data pre-processing is required before use. There is no so-called 'standard' process for data preprocessing, it depends on the different tasks and data set attributes. The text preprocessing steps taken in this paper is shown as following:

1. Word tokenization: After removing the punctuation marks of the text, we use the package in Python named 'tokenize' to have the original text document decompose into a list, in which vocabulary is separated by a space.
2. Stopwords removing: Generally speaking, stop words are roughly divided into two categories. One type is the functional words contained in the human language. These functional words are extremely common. Compared with other words, functional words have no actual meaning, such as 'the', 'is', 'at', 'which, 'on'. Another type of words includes a modal verb, such as 'want', 'could', 'can'. These words are widely used, but there is not enough practical meaning for text understanding or classification. In this paper, we download the stop words list from a package in Python named '**NLTK**'.
3. Finding word stems: Stemming is the removal of the plural of some nouns, the

removal of different tenses of verbs, etc.

4. Lemmatization: Lemmatization is to remove the affix of the word, extract the main part of the word, and restore the prototype of the word. For example, the word "cars" is reduced to the word "car", and the word "ate" is "eat".

## 4.3 Word embedding and similarity analysis

There are two ways to transform the unstructured words data to a structured dataset, which are one hot representation and distributed representation respectively. Word embedding is a distributed expression. The idea is to map each word to a shorter word vector through training. The mapping must satisfy two conditions: (1) This mapping is injective; (2) The vector after the mapping will not lose the information contained in the previous vector. Word2Vec is a simplified neural network model. The input is a One-Hot Vector, Hidden Layer has no activation function, which is a linear unit. The output layer dimension is the same as the input layer dimension, using softmax regression. After the model is trained, what we need is the weight matrix of the hidden layer that the model learned through the training data. We use the Skip-Gram model as the training model, which performs better in large corpora (2008-2018 compliance reports data). The input of Skip-Gram is specific to a certain word, and the output is the context vector corresponding to the word. That is to predict the context through the given input word.

# 5. Similarity analysis results and discussions

In this section, we first discuss the environmental pollution risks that may occur during the UOG production process from previous literature. Based on the discussion results and the statistical analysis of **Section 3**, we traversed the original corpus and selected keywords of "Location", "Operation" and "Contaminant" respectively. Finally, this article conducts a keyword similarity analysis to obtain the internal mechanism and general laws of violations.

## 5.1 Keywords selection

First, we build the "**Environmental Health & Safety**" original corpora. The top 20 frequently words list of type "**Environmental Health & Safety**" category in *Inspect Comment* is as shown in **Table.1**. Break down the factors of violation accidents, we divide these keywords in this article into "Location", "Contaminant" and "Operation" three categories. "Location" implicates this kind of words can fully represent the place where the violation occurred or where be polluted. "Contaminant" represents the materials that caused pollution, while "Operation" indicates the behavior and operation processed during UOG production.

**Table 1.** The top 20 words in "Environmental Health & Safety" category of violation compliance

| Violation type | (Word, frequency) |
|---|---|
| Environmental Health & Safety | ('area', 1779), ('surface', 1802), ('product', 1817), ('gas', 1857), ('tank', 2003), ('fluid', 2042), ('release', 2180), ('depart', 2265), ('report', 2451), ('conduct', 2598), ('spill', 2675), ('time', 2731), ('operate', 2842), ('water', 3250), ('contain', 3602), ('pad', 4142), ('violate', 4829), ('site', 5326), ('well', 6515), ('inspect', 7697) |

The environmental pollution caused by UOG extraction can be roughly divided into three categories: water, soil, and atmospheric pollution. Water pollution is mainly caused by the leakage of sewage during transportation, water treatment, and fracturing. Geological hazards are mainly caused by drilling and fracturing, and even triggering earthquakes, or pollution incidents caused by chemical substances leaking into the soil. Air pollution mainly occurs during venting or unintended leaks, due to the residual methane gas leaking from the shaft, as well as the greenhouse gas emissions caused by transportation and production. Among them, water pollution is the most threatening type of pollution to human health and the environment, and because it is easier to track than the other two sources of pollution, it is currently the most important focus of UOG environmental regulations.

There are several sites and transportation methods. First of all, the control center engaged in water transportation and material transportation and mining plans. After the drilling address and scale were determined, water was passed through the pipeline, and other fracturing auxiliary materials were sent to the site by chemical trucks to perform mixing work through the blender. After the fracturing is completed, the mixed proppant is pressed into the well, the purpose is to support the fractures formed in the horizontal well after fracturing, and then the liquid mixed with UOG and fracturing fluid is extracted through the well and transported to the pipeline The processing center on the well separate the oil and gas from the return fluid. The separated flowback fluid can be reused in the water management center to be mixed into fracturing fluid for storage or discharged after treatment.

According to the previous literature (Torres et al., 2016; Werner et al., 2015) and the corpus of violations from 2008 to 2018, we compiled the following figure to initially show the relationship between environment & health impact media, violations key factors, high-risk activities and possible pollution violations during UOG production as shown in **Figure 4**.

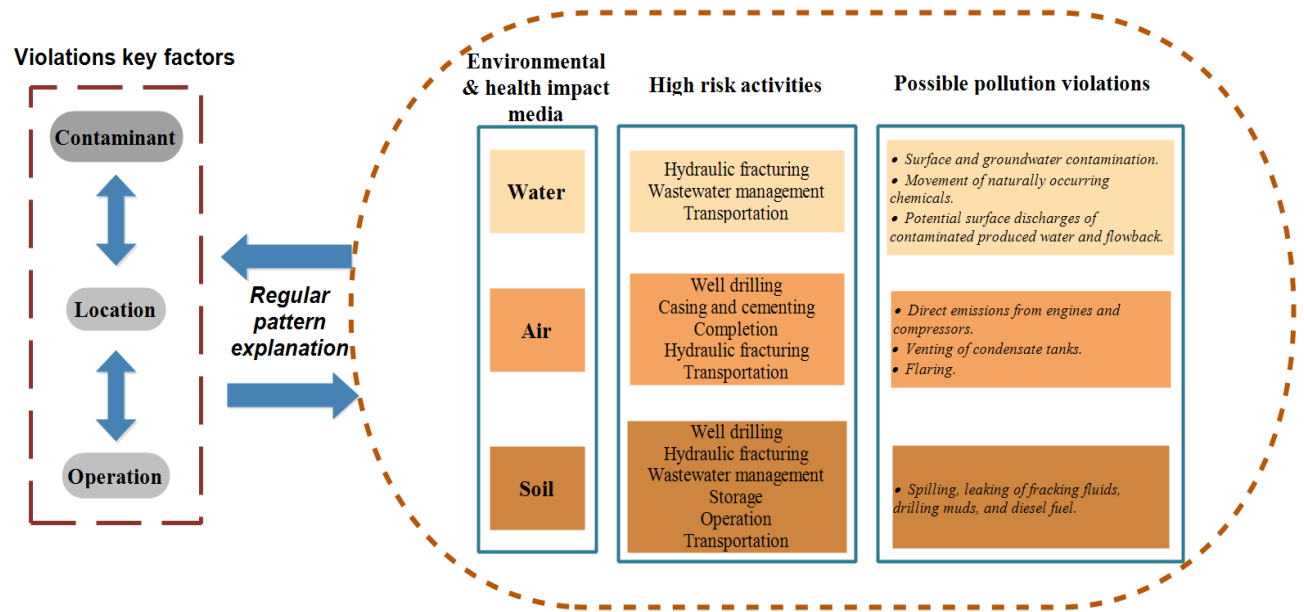

**Figure 4.** Keywords selection process

Ultimately, we determined the violations accident keywords as shown in **Table 2**. Among them: (1)"Contaminant" includes 12 keywords and the types of contaminants can be roughly divided into wastewater, waste gas, fuel, and solid waste. The pollutants that frequently occur in pollution accidents are "gas" and "methane", followed by sewage "brine", "flowback" and "drainage", followed by oil pollutants "oil" and "fuel", and finally solid pollutants including "mud" and "silt"; (2) while "Location" contains 12 keywords. Among the most mentioned are the broad locations, such as "well" and "pad". The more specific locations are "tank" first, followed by "truck" and "road" involving transportation, and then on-site savings and production equipment, such as "barrel", "sump", "mat", etc..; (3) "Operation" contains 11 keywords. The most accident-related operations that appear in the violation review are "spill", followed by "drill", "vent" and "excavate". Judging from the frequency of occurrence of such words, the phases with the highest number of violations include drilling, hydraulic fracturing, and waste disposal.

**Table 2.** Keywords categories of "Environmental Health & Safety" compliance

| Categories of corpus | Top frequent Words(frequency) |
|---|---|
| Contaminant | ('gas', 1857), ('brine', 615), ('sediment', 501), ('oil', 493), ('flowback', 404), ('mud', 400), ('diesel', 337), ('fuel', 272), ('methane', 222), ('drain', 132), ('silt', 115), ('drainage', 107) |
| Location | ('well', 6515), ('pad', 4142), ('tank', 2003), ('road', 773), ('truck', 741),('ground', 688), ('barrel', 505), ('puddle', 369), ('pit', 286), ('sump', 263), ('mat', 245), ('impound', 159) |
| Operation | ('spill', 2675), ('drill', 1314), ('vent', 1027),('excavate', 934), ('case', 572), ('frack', 570), ('erosion', 374), ('leak', 355), ('combust', 319), ('dispose', 304) |

## 5.2 Similarity results and discussions

According to the analysis of frequent violations in the basic analysis as well as the production process associated with the violation risk analysis mentioned above, we choose necessary and representative terminologies traversing inspection original corpus as shown in **Table 2**, which appeared more than 30 times of total 5214 original word corpus. and then calculate similarity in pairs, which is evaluated by calculating the cosine of the angle between the two words embedding as **Eq.1**. It is most commonly used in high-dimensional positive spaces, given two words embedding as A and B:

$$similarity = \cos(\theta) = \frac{A \cdot B}{\|A\|\|B\|} = \frac{\sum_{i=1}^{n} A_i \times B_i}{\sqrt{\sum_{i=1}^{n}(A_i)^2} \times \sqrt{\sum_{i=1}^{n}(B_i)^2}} \qquad (1)$$

Cosine similarity value is between $[0,1]$, the higher the cosine similarity between two words embedding, the more relevant they are in the text, the more frequently they appear together.

According to three categories of terminologies, which are aforementioned "location", "contaminants" and "operation" respectively, the pair similarity analysis is conducted as shown in **Table 3-5**. We will get the correlation between the three categories of keywords and interesting observations from these three pairwise similarity tables.

From the perspective of **Location & Contaminant** about UOG environmental health violations as shown in **Table 3**. We can get some preliminary conclusions: (1) the closest relationship is "fuel" to "mat", followed by "silt" to "pit" and "diesel" to "mat". From the perspective of pollutants, the pollution locations of solid wastes are mainly in "pit", "ground" and "mat". Sewage type such as "flowback" ", "brine", "drainage", their pollution locations are mainly in "tank", "truck", "blender", "road", "impound", "pad" and "sump". The pollution locations of "gas" and "methane" are mainly in "well". The pollution sites of "oil", "fuel" and "diesel" are mainly in "barrel", " "road", "blender", "truck", "puddle" and "mat"; (2) the pollution range of sewage is the largest, the leakage place is mainly in the storage tank, the storage pool as well as wastewater management equipment, and there may also be a risk of sewage leakage during transportation； (3) the two main types of pollution accidents involving oil pollutants include vehicle leakage during transportation and oil leakage from certain equipment during operation； (4) the pollution sources of exhaust pollutants are mainly in wells and storage equipment.

From the perspective of the relationship between contaminant and operation, as shown in **Table 4**, there are some preliminary findings: (1) the most closely related relationships include: "waste" and "dispose ", "methane" and "combust", "sediment" and "erosion", "silt" and "erosion" Among them, the keyword of the violation incident that is most closely related to the pollutant is "erosion", and the keyword of the pollutant most closely associated with the violation operation is "methane", followed by "fuel"; (2) due to the actions of "dispose", "erosion" and "drill", the illegal discharge of solid waste "waste", "sediment", "silt" and "mud" will be caused; (3) due to the operation of "leak", "frack", "erosion" and "excavate", violations of sewage pollutants will be caused; (4) violations of oil wastes are mainly related to "drill", "leak" and "spill"; (5) among them, "combust", "vent", "case", "cement" operation is the most likely to cause illegal emission accidents of exhaust gas.

From the relationship of operation and location as **Table 5**, we find that the most closely related relations are "pit" and "erosion", "sump" and "dispose", "barrel" and "leak". In this kind of pairs of keyword connections, the lack of contaminants makes the appearance of certain keyword pairs meaningless. Therefore, we will further analyze the results of the similarity analysis based on the relationship from "Location" to "Operation" to "Contaminant".

**Table 3**. Contaminant and location keywords similarity.

| Location \ Contaminant | waste | brine | gas | methane | oil | fuel | flowback | drainage | sediment | silt | mud | diesel |
|---|---|---|---|---|---|---|---|---|---|---|---|---|
| pit | **0.458** | 0.275 | 0.078 | 0.073 | 0.173 | 0.202 | 0.239 | 0.300 | **0.415** | <u>0.473</u> | 0.316 | 0.233 |
| road | 0.264 | 0.163 | 0.016 | 0.116 | 0.214 | 0.365 | 0.217 | <u>0.460</u> | 0.310 | 0.315 | 0.241 | 0.312 |
| tank | 0.172 | <u>**0.406**</u> | 0.169 | 0.184 | 0.134 | 0.223 | 0.304 | 0.246 | 0.199 | 0.200 | 0.221 | 0.150 |
| truck | 0.243 | 0.163 | 0.090 | 0.123 | 0.104 | 0.318 | <u>0.320</u> | 0.251 | 0.250 | 0.172 | 0.265 | 0.277 |
| blender | 0.309 | 0.332 | 0.122 | 0.239 | 0.109 | 0.351 | <u>0.390</u> | 0.249 | 0.135 | 0.196 | 0.242 | 0.307 |
| impound | 0.221 | 0.277 | 0.047 | 0.078 | 0.010 | 0.161 | 0.298 | <u>0.384</u> | 0.280 | 0.326 | 0.182 | 0.167 |
| well | 0.107 | 0.198 | <u>**0.347**</u> | 0.148 | 0.202 | 0.083 | 0.221 | 0.155 | 0.125 | 0.177 | 0.231 | 0.148 |
| pad | 0.112 | 0.215 | 0.113 | 0.029 | 0.140 | 0.250 | 0.330 | <u>0.370</u> | 0.185 | 0.180 | 0.292 | 0.293 |
| ground | <u>0.390</u> | 0.190 | 0.166 | 0.256 | 0.256 | 0.357 | 0.337 | 0.304 | 0.152 | 0.184 | 0.367 | 0.359 |
| barrel | 0.216 | 0.270 | 0.035 | 0.153 | **0.258** | 0.306 | 0.291 | 0.226 | 0.166 | 0.154 | <u>0.379</u> | 0.330 |
| puddle | 0.201 | 0.203 | 0.096 | 0.216 | 0.019 | <u>0.306</u> | 0.305 | 0.205 | 0.247 | 0.108 | 0.133 | 0.232 |
| sump | 0.202 | 0.274 | 0.045 | 0.110 | 0.079 | 0.176 | 0.205 | <u>**0.476**</u> | 0.309 | 0.329 | 0.256 | 0.233 |
| mat | 0.269 | 0.199 | 0.165 | 0.109 | 0.117 | <u>**0.512**</u> | 0.155 | 0.158 | 0.210 | 0.290 | **0.382** | **0.483** |

Note: "_" means the number with the largest value in each row. **Bold** means the largest number in each column.

**Table 4.** Contaminant and operation keywords similarity.

| Operation\Contaminant | case | erosion | drill | spill | leak | cement | vent | dispose | frack | excavate | combust |
|---|---|---|---|---|---|---|---|---|---|---|---|
| waste | 0.125 | 0.329 | 0.212 | 0.223 | 0.154 | 0.132 | 0.104 | **0.656** | 0.167 | 0.243 | 0.060 |
| brine | 0.135 | 0.262 | 0.179 | 0.353 | 0.399 | 0.082 | 0.203 | 0.246 | 0.220 | 0.248 | 0.274 |
| gas | 0.388 | 0.136 | 0.181 | 0.102 | 0.280 | 0.249 | 0.372 | 0.024 | 0.148 | 0.130 | 0.558 |
| methane | **0.489** | 0.212 | 0.231 | 0.137 | 0.197 | **0.379** | **0.545** | 0.000 | 0.183 | 0.106 | **0.692** |
| oil | 0.246 | 0.156 | 0.357 | 0.275 | 0.285 | 0.192 | 0.201 | 0.110 | 0.142 | 0.099 | 0.174 |
| fuel | 0.097 | 0.196 | 0.279 | 0.400 | **0.416** | 0.173 | 0.175 | 0.157 | 0.154 | 0.161 | 0.167 |
| flowback | 0.096 | 0.165 | 0.391 | 0.298 | 0.125 | 0.155 | 0.126 | 0.151 | **0.452** | 0.209 | 0.110 |
| drainage | 0.107 | 0.571 | 0.000 | 0.238 | 0.281 | 0.071 | 0.151 | 0.256 | 0.228 | **0.314** | 0.310 |
| sediment | 0.089 | 0.659 | 0.119 | 0.142 | 0.162 | 0.062 | 0.109 | 0.140 | 0.109 | 0.230 | 0.205 |
| silt | 0.180 | **0.679** | 0.148 | 0.120 | 0.125 | 0.186 | 0.171 | 0.282 | 0.169 | 0.236 | 0.219 |
| mud | 0.199 | 0.308 | **0.460** | 0.280 | 0.214 | 0.181 | 0.280 | 0.065 | 0.289 | 0.216 | 0.264 |
| diesel | 0.051 | 0.110 | 0.337 | **0.369** | 0.342 | 0.077 | 0.166 | 0.165 | 0.143 | 0.198 | 0.154 |

Note: "_" means the number with the largest value in each row. **Bold** means the largest number in each column.

**Table 5.** Operation and location keywords similarity.

| Operation\Location | case | erosion | drill | spill | leak | cement | vent | dispose | frack | excavate | combust |
|---|---|---|---|---|---|---|---|---|---|---|---|
| pit | 0.141 | 0.439 | 0.243 | 0.280 | **0.395** | 0.184 | 0.101 | 0.314 | 0.191 | 0.274 | 0.114 |
| road | 0.176 | **0.437** | 0.216 | 0.230 | 0.169 | 0.048 | 0.166 | 0.249 | 0.151 | 0.175 | 0.022 |
| tank | 0.149 | 0.235 | 0.134 | 0.250 | 0.256 | 0.133 | **0.352** | 0.254 | 0.325 | 0.160 | 0.206 |
| truck | 0.066 | 0.161 | 0.166 | 0.267 | 0.284 | 0.057 | 0.169 | 0.220 | 0.226 | 0.269 | 0.075 |
| blender | **0.236** | 0.180 | 0.226 | 0.366 | 0.334 | **0.311** | 0.203 | 0.348 | **0.467** | 0.305 | 0.107 |
| impound | 0.111 | 0.404 | 0.262 | 0.159 | 0.284 | 0.059 | 0.000 | 0.267 | 0.288 | 0.215 | 0.029 |
| well | 0.219 | 0.096 | **0.345** | 0.096 | 0.196 | 0.216 | 0.324 | 0.115 | 0.188 | 0.195 | **0.285** |
| pad | 0.133 | 0.245 | 0.201 | 0.290 | 0.228 | 0.136 | 0.164 | 0.252 | 0.235 | 0.318 | 0.161 |
| ground | 0.206 | 0.247 | 0.204 | 0.286 | 0.253 | 0.156 | 0.202 | 0.330 | 0.128 | 0.228 | 0.160 |
| barrel | 0.115 | 0.133 | 0.195 | **0.422** | 0.220 | 0.108 | 0.103 | 0.267 | 0.282 | 0.255 | 0.102 |
| puddle | 0.199 | 0.197 | 0.126 | 0.269 | 0.254 | 0.140 | 0.263 | 0.189 | 0.223 | 0.289 | 0.224 |
| sump | 0.060 | 0.369 | 0.073 | 0.193 | 0.193 | -0.027 | 0.201 | **0.455** | 0.313 | **0.430** | 0.148 |
| mat | 0.187 | 0.193 | 0.258 | 0.268 | 0.270 | 0.241 | 0.215 | 0.190 | 0.200 | 0.249 | 0.277 |

Note: "_" means the number with the largest value in each row. **Bold** means the largest number in each column.

After organizing, we obtain the relationship from "Location" to "Operation" to "Contaminant" as shown in **Table 6**. Among them, "Location-Contaminant" means contaminants that have a higher relevance (upper quartile) appear in the location, "Operation-Contaminant" denotes high-risk (upper quartile) contaminants directly caused by certain operations. We can see the high-risk operation behaviors in different locations and different equipment and the corresponding pollutants. Therefore, we hold following points of view: (1) among the keywords that represent deposits, the most frequent types of pollutants are sediment and sewage, which are usually caused by disposing, erosion, and overflow of deposits; (2) among the location keywords, the most frequent types of pollutants are oil, sewage, and silt. The main reason is the diesel spill caused by the failure of the transportation truck, the overflow during sewage transportation; (3) the drilling process and exhaust process are the most likely to occur methane illegal emission accident.

**Table 6.** The relationship of "Location" to "Operation" to "Contaminant".

| Location-Contaminant | Location | Operation | Operation-Contaminant |
|---|---|---|---|
| Waste, sediment, silt | Pit | Erosion | Drainage, sediment, silt |
| | | Leak | Brine, fuel, diesel |
| | | Dispose | Waste, drainage, silt |
| Fuel, drainage, silt | Road | Erosion | Drainage, sediment, silt |
| | | Spill | Brine, fuel, diesel |
| | | Dispose | Waste, drainage, silt |
| Brine, flowback, drainage | Tank | Vent | Gas, methane, mud |
| | | Leak | Brine, fuel, diesel |
| | | Frack | Flowback, drainage, mud |
| Flowback, fuel, diesel | Truck | Leak | Brine, fuel, diesel |
| | | Spill | Brine, fuel, diesel |
| | | Excavate | Waste, brine, drainage |
| Flowback, brine, fuel | Blender | Spill | Brine, fuel, diesel |
| | | Dispose | Waste, drainage, silt |
| | | Frack | Flowback, drainage, mud |
| Drainage, flowback, silt | Impound | Erosion | Drainage, sediment, silt |
| | | Leak | Brine, fuel, diesel |
| | | Frack | Flowback, drainage, mud |
| Gas, flowback, mud | Well | Drill | Oil, flowback, mud |
| | | Combust | Gas, methane, drainage |
| | | Vent | Gas, methane, mud |
| Drainage, fuel, drainage | Pad | Excavate | Waste, brine, drainage |
| | | Spill | Brine, fuel, diesel |
| | | Dispose | Waste, drainage, silt |
| Waste, mud, diesel | Ground | Dispose | Waste, drainage, silt |
| | | Spill | Brine, fuel, diesel |
| | | Leak | Brine, fuel, diesel |
| Fuel, mud, diesel | Barrel | Spill | Brine, fuel, diesel |
| | | Dispose | Waste, drainage, silt |
| | | Frack | Flowback, drainage, mud |
| Fuel, flowback, drainage | Puddle | Excavate | Waste, brine, drainage |

|  |  | Spill | Brine, fuel, diesel |
|  |  | Vent | Gas, methane, mud |
| Drainage, sediment, silt | Sump | Dispose | Waste, drainage, silt |
|  |  | Erosion | Drainage, sediment, silt |
|  |  | Excavate | Waste, brine, drainage |
| Fuel, mud, diesel | Mat | Leak | Brine, fuel, diesel |
|  |  | Spill | Brine, fuel, diesel |
|  |  | Combust | Gas, methane, drainage |

# 6. Conclusion and future work

The purpose of this article is to use text mining and natural language processing technologies to analyze the text data of UOG violations, which are difficult to handle manually, and discover the general laws of environmental violations, ultimately provide corresponding environmental supervision strategies to government and practitioners. The data source comes from the Pennsylvania Department of Environmental Protection, we select the UOG violation compliance report from 2008/01/01 to 2018/12/31. According to the violation regulations and previous literature research, this article decomposes the original thesaurus of UOG violation original corpus into three types of keywords, namely "operation", "location" and "contaminant". Then, using the Skip-Gram model, the text data was converted into word embedding, and the cosine similarity between keywords were used to find the high-frequency violations and pollutants corresponding to different locations in environmental and health safety violations.

After analyzing the statistics of violation historical data and the similarity of word vectors, this article summarizes the following main findings: (1) the pollution range of sewage is the largest, the leakage place is mainly in the storage tank, the storage pool as well as wastewater management equipment, and there may also be a risk of sewage leakage during transportation； (2) the two main types of pollution accidents involving oil pollutants include vehicle leakage during transportation and oil leakage from certain equipment during operation； (3) the pollution sources of exhaust pollutants are mainly in wells and storage equipment; (4) among the keywords that represent deposits, the most frequent types of pollutants are sediment and sewage, which are usually caused by damage, corrosion, and overflow of deposits; (5) among the location keywords, the most frequent types of pollutants are oil, sewage, and silt. The main reason is the diesel spill caused by the failure of the transportation truck, the overflow during sewage transportation; (6) the drilling process and exhaust process are the most likely to occur methane illegal emission accident.

The contributions of this paper are mainly about three aspects: (1) To the best of our knowledge, this article is the first one to apply natural language processing technology to the field of environmental health management of UOG, which can inspire follow-up research to further discuss this technology on its industrial supervision. (2) For the first time, this article processed the environmental violation review data within 10 years (from 2008 to 2018), and used word embedding similarity analysis to show the association between violation risk, accident location and operation behavior. (3) From the perspective of practitioners, the correlation results obtained from the review of accident data based on environmental violations can be applied to actual site management, which can remind practitioners to pay attention to certain frequently occurring problems on the operation site to establish a better monitoring system.

However, some limitations are existing: (1) we use word embedding instead of term embedding or document embedding, which can be further analyzed in follow-up work; (2) we have not solved the problem of polysemy, and we can consider improving it in the follow-up work; (3) the keywords selected in this article are selected by the author team after personal consideration and discussion. Therefore, we plan to adopt an unsupervised method to select keywords in future work. (4) unfortunately, due to the limitations of the original data, more detailed key factors identification cannot be performed. Therefore, the next step of this research is to consider the integration of violations database of text data.

# Acknowledgments

This research work was partly supported by the National Natural Science Foundation of China under Grant Nos. 71774130 and 71988101; the Fundamental Research Funds for the Central Universities under Grant No. xpt012020022.

# Conflicts of interest

The authors declare that there are no conflicts of interest regarding the publication of this study.

# References:

Abualfaraj, N., M. S. Olson, P. L. Gurian, A. De Roos, and C. A. Gross-Davis, 2016, Statistical analysis of compliance violations for natural gas wells in Pennsylvania: Energy Policy, v. 97, p. 421-428.

Butkovskyi, A., G. Cirkel, E. Bozileva, H. Bruning, A. P. Van Wezel, and H. H. M. Rijnaarts, 2019, Estimation of the water cycle related to shale gas production under high data uncertainties: Dutch perspective: Journal of Environmental Management, v. 231, p. 483-493.

Casey, J. A., D. E. Goin, K. E. Rudolph, B. S. Schwartz, D. Mercer, H. Elser, E. A. Eisen, and R. Morello-Frosch, 2019, Unconventional natural gas development and adverse birth outcomes in Pennsylvania: The potential mediating role of antenatal anxiety and depression: Environ Res, v. 177, p. 108598.

Castro-Alvarez, F., P. Marsters, D. Ponce De León Barido, and D. M. Kammen, 2018, Sustainability lessons from shale development in the United States for Mexico and other emerging unconventional oil and gas developers: Renewable and Sustainable Energy Reviews, v. 82, p. 1320-1332.

Centner, T. J., 2016, Reducing pollution at five critical points of shale gas production: Strategies and institutional responses: Energy Policy, v. 94, p. 40-46.

Centner, T. J., and L. K. O'Connell, 2014, Unfinished business in the regulation of shale gas production in the United States: Science of The Total Environment, v. 476-477, p. 359-367.

Chen, H., and K. E. Carter, 2016, Water usage for natural gas production through hydraulic fracturing in the United States from 2008 to 2014: Journal of Environmental Management, v. 170, p. 152-159.

Chen, Y., J. Li, H. Lu, and J. Xia, 2020, Tradeoffs in water and carbon footprints of shale gas, natural gas, and coal in China: Fuel, v. 263, p. 116778.

Cronshaw, I., and R. Q. Grafton, 2016, Economic benefits, external costs and the regulation of unconventional gas in the United States: Energy Policy, v. 98, p. 180-186.

Davis, R. J., and C. Sims, 2019, Frack to the future: What enticed small firms to enter the natural gas market during the hydraulic fracturing boom? Energy Economics, v. 81, p. 960-973.

Deziel, N. C., E. Brokovich, I. Grotto, C. J. Clark, Z. Barnett-Itzhaki, D. Broday, and K. Agay-Shay, 2020, Unconventional oil and gas development and health outcomes: A scoping review of the epidemiological research: Environ Res, v. 182, p. 109124.

Guo, M., Y. Xu, and Y. D. Chen, 2017, Catching environmental noncompliance in shale gas development in China and the United States: Resources, Conservation & Recycling, v. 121, p. 73-81.

Guo, M., Y. Xu, and Y. D. Chen, 2019, Environmental enforcement and compliance in Pennsylvania's Marcellus shale gas development: Resources, Conservation & Recycling, v. 144, p. 24-31.

Harleman, M., and J. G. Weber, 2017, Natural resource ownership, financial gains, and governance: The case of unconventional gas development in the UK and the US: Energy Policy, v. 111, p. 281-296.

Hernández-Espriú, A., B. Wolaver, S. Arciniega-Esparza, B. R. Scanlon, M. H. Young, J. Nicot, S. Macías-Medrano, and J. A. Breña-Naranjo, 2019, A screening approach to improve water management practices in undeveloped shale plays, with application to the transboundary Eagle Ford Formation in northeast Mexico: Journal of Environmental Management, v. 236, p. 146-162.

Langlois, L. A., P. J. Drohan, and M. C. Brittingham, 2017, Linear infrastructure drives habitat conversion and forest fragmentation associated with Marcellus shale gas development in a forested landscape: Journal of Environmental Management, v. 197, p. 167-176.

Maloney, K. O., S. Baruch-Mordo, L. A. Patterson, J. Nicot, S. A. Entrekin, J. E. Fargione, J. M. Kiesecker, K. E. Konschnik, J. N. Ryan, A. M. Trainor, J. E. Saiers, and H. J. Wiseman, 2017, Unconventional oil and gas spills: Materials, volumes, and risks to surface waters in four states of the U.S.: Science of The Total Environment, v. 581-582, p. 369-377.

Manda, A. K., J. L. Heath, W. A. Klein, M. T. Griffin, and B. E. Montz, 2014, Evolution of multi-well pad development and influence of well pads on environmental violations and wastewater volumes in the Marcellus shale (USA): Journal of Environmental Management, v. 142, p. 36-45.

Oke, D., R. Mukherjee, D. Sengupta, T. Majozi, and M. El-Halwagi, 2020, On the optimization of water-energy nexus in shale gas network under price uncertainties: Energy, v. 203, p. 117770.

Rahm, B. G., S. Vedachalam, L. R. Bertoia, D. Mehta, V. S. Vanka, and S. J. Riha, 2015, Shale gas operator violations in the Marcellus and what they tell us about water resource risks: Energy Policy, v. 82, p. 1-11.

Solarin, S. A., L. A. Gil-Alana, and C. Lafuente, 2020, An investigation of long range reliance on shale oil and shale gas production in the U.S. market: Energy, v. 195, p. 116933.

Sun, Y., D. Wang, D. C. W. Tsang, L. Wang, Y. S. Ok, and Y. Feng, 2019, A critical review of risks, characteristics, and treatment strategies for potentially toxic elements in wastewater from shale gas extraction: Environment International, v. 125, p. 452-469.

Torres, L., O. P. Yadav, and E. Khan, 2016, A review on risk assessment techniques for hydraulic fracturing water and produced water management implemented in onshore unconventional oil and gas production: Science of The Total Environment, v. 539, p. 478-493.

Wang, Q., and F. Jiang, 2019, Integrating linear and nonlinear forecasting techniques based on grey theory and artificial intelligence to forecast shale gas monthly production in Pennsylvania and Texas of the United States: Energy, v. 178, p. 781-803.

Wang, Q., and L. Zhan, 2019, Assessing the sustainability of the shale gas industry by combining DPSIRM model and RAGA-PP techniques: An empirical analysis of Sichuan and Chongqing, China: Energy, v. 176, p. 353-364.

Watson, T. L., and S. Bachu, 2009, Evaluation of the Potential for Gas and CO2 Leakage Along Wellbores: SPE Drilling & Completion, v. 24, p. 115-126.

Werner, A. K., S. Vink, K. Watt, and P. Jagals, 2015, Environmental health impacts of unconventional natural gas development: A review of the current strength of evidence: Science of The Total Environment, v. 505, p. 1127-1141.

Wisen, J., R. Chesnaux, G. Wendling, J. Werring, F. Barbecot, and P. Baudron, 2019, Assessing the potential of cross-contamination from oil and gas hydraulic fracturing: A case study in northeastern British Columbia, Canada: Journal of Environmental Management, v. 246, p. 275-282.

Zirogiannis, N., J. Alcorn, J. Rupp, S. Carley, and J. D. Graham, 2016, State regulation of unconventional gas development in the U.S.: An empirical evaluation: Energy Research & Social Science, v. 11, p. 142-154.